\documentclass[conference]{IEEEtran}
\IEEEoverridecommandlockouts
\usepackage{cite}
\usepackage{amsmath,amssymb,amsfonts}
\usepackage{algorithmic}
\usepackage{graphicx}
\usepackage{textcomp}
\usepackage{placeins}
\usepackage{array}
\usepackage[table]{xcolor}

\usepackage{graphicx}
\usepackage{caption}
\usepackage{pgfplots}
\usepackage{tikz}
\usetikzlibrary{shapes.geometric, arrows}
\usepackage{tkz-graph}
\usepackage{algorithm}
\usepackage{hyperref}
\usepackage{cleveref}
\usepackage{comment}

\def\BibTeX{{\rm B\kern-.05em{\sc i\kern-.025em b}\kern-.08em
    T\kern-.1667em\lower.7ex\hbox{E}\kern-.125emX}}
    
\begin{document}

\title{\title{FUSION: A Flexible Unified Simulator for Intelligent Optical Networking \\
\thanks{This paper was partially supported by NSF project award \#2008530.}
}
}

\author{
    \IEEEauthorblockN{\hspace*{-2.9cm} Ryan McCann$^{\dagger,*}$}
    \IEEEauthorblockA{\hspace*{-2.9cm} \textit{ryan\_mccann@student.uml.edu}}
    \and
    \IEEEauthorblockN{Arash Rezaee$^{\dagger,*}$}
    \IEEEauthorblockA{\textit{arash\_rezaee@student.uml.edu}}
    \and
    \IEEEauthorblockN{Vinod M. Vokkarane$^\dagger$}
    \IEEEauthorblockA{\textit{vinod\_vokkarane@uml.edu}}
    \and
    \makebox[\linewidth]{\textit{$^\dagger$Electrical and Computer Engineering Department, University of Massachusetts Lowell, United States}}
    \and
    \makebox[\linewidth]{\textit{$^*$Both authors contributed equally to this work.}}
}

\maketitle

\begin{abstract}
The increasing demand for flexible and efficient optical networks has led to the development of Software-Defined Elastic Optical Networks (SD-EONs). These networks leverage the programmability of Software-Defined Networking (SDN) and the adaptability of Elastic Optical Networks (EONs) to optimize network performance under dynamic traffic conditions. However, existing simulation tools often fall short in terms of transparency, flexibility, and advanced functionality, limiting their utility in cutting-edge research. In this paper, we present a Flexible Unified Simulator for Intelligent Optical Networking (FUSION), a fully open-source simulator designed to address these limitations and provide a comprehensive platform for SD-EON research. FUSION integrates traditional routing and spectrum assignment algorithms with advanced machine learning and reinforcement learning techniques, including support for the Stable Baselines 3 library. The simulator also offers robust unit testing, a fully functional Graphical User Interface (GUI), and extensive documentation to ensure usability and reliability. Performance evaluations demonstrate the effectiveness of FUSION in modeling complex network scenarios, showcasing its potential as a powerful tool for advancing SD-EON research.
\end{abstract}


\section{Introduction}

The increasing global demand for internet services has driven the need for advanced communication networks that can adapt to growing data volumes and dynamic traffic patterns \cite{Cisco2023InternetReport}. Elastic Optical Networks (EONs) have emerged as a key technology, offering flexible bandwidth allocation and improved spectral efficiency to meet these demands. Optical networking, which forms the backbone of modern telecommunications, traditionally relies on fixed Wavelength Division Multiplexing (WDM) systems. However, the rigid nature of WDM has led to the development of EONs, which provide a more adaptable approach to spectrum management \cite{Gerstel2012EONs}. With the advent of Software Defined Networking (SDN), the capabilities of EONs have been enhanced, giving rise to Software Defined Elastic Optical Networks (SD-EONs), which enable more dynamic and efficient network configurations \cite{Nisar2020SDN}.

Despite the importance of SD-EONs, the research community lacks a comprehensive, open-source simulator for rigorous testing and development in this field. Existing simulators often fall short in areas such as transparency, flexibility, and advanced functionality. Many remain closed-source, limiting their adoption and trustworthiness. This gap highlights the need for a high-quality, open-source SD-EON simulator.

The primary objective of this paper is to present the development of a fully open-source software defined elastic optical networking simulator. This simulator is designed to meet current research needs and offer features lacking in other tools, including thorough unit testing, detailed documentation, and integration with modern machine learning frameworks such as Reinforcement Learning (RL), with support for the Stable Baselines 3 library \cite{scikit-learn, stable-baselines3}. Additionally, the simulator includes a fully functional Graphical User Interface (GUI) and advanced output capabilities, such as data export in multiple formats. By maintaining a rigorous software development pipeline and fostering community contributions, this simulator aims to set a new standard in SD-EON research.

The paper is structured as follows: Section \ref{sec:literature_review} reviews existing optical network simulators. Section \ref{sec:system_architecture} outlines the system architecture of the proposed simulator. Section \ref{sec:methodology} details the simulator's functionality and validation setups. Section \ref{sec:performance_evaluation} presents the performance evaluation. Finally, Section \ref{sec:conclusion} concludes the paper and discusses future research directions.

\section{Literature Review}
\label{sec:literature_review}

Software Defined Optical Networks (SDONs) represent a significant evolution in the field of optical networking, combining the flexibility of Elastic Optical Networks (EONs) with the programmability of Software Defined Networking (SDN). EONs enable dynamic spectrum allocation, optimizing the use of available resources to accommodate varying traffic demands \cite{Gerstel2012EONs}. When integrated with SDN, these networks benefit from centralized control, allowing for more efficient network management and adaptability in response to real-time changes \cite{Nisar2020SDN}.

Several simulators have been developed to model and evaluate the performance of SDONs, each offering distinct features and functionalities. These tools are crucial for advancing research in the field, enabling the exploration of new algorithms and network configurations under controlled conditions. Among these, Optical RL-Gym stands out for its integration with the Stable Baselines 3 library, providing a comprehensive platform for RL experiments in optical networks \cite{optical-rl-gym}. It offers a variety of custom environments tailored for Deep RL, such as DeepRMSA, which is an adapted version of a previously published routing and spectrum assignment (RSA) algorithm \cite{deep-rmsa}. Optical RL-Gym also includes documentation and tests that facilitate the creation and use of custom environments, although the level of detail in these resources varies. The simulator is fully open-source, making it accessible for modification and extension by researchers.

OMNeT++ is a modular discrete-event simulation framework widely used in networking research, including optical networks \cite{omnetpp}. Its strength lies in its extensibility and the large ecosystem of models like INET for various network types. OMNeT++ offers a user-friendly GUI for simulation setup and analysis, making it easier to configure network scenarios and monitor performance metrics. While the framework is open-source, some specialized modules require licensing, which can limit accessibility. Despite its steep learning curve, OMNeT++ is popular for its flexibility and strong community support.

SimTON, another prominent simulator, focuses on providing highly customizable simulation parameters, particularly for physical layer aspects like input power, gain factor, and the residual dispersion parameter \cite{simton}. It features a robust GUI that supports topology visualization and allows users to interact with nodes and links to access detailed information. SimTON’s detailed physical layer assumptions and the ease of adjusting optical components such as switches, amplifiers, and fibers make it a powerful tool for simulating complex optical networks.

SimEON offers a different set of capabilities, particularly in modeling physical impairments, dynamic traffic, and energy consumption within optical networks \cite{simeon}. It supports regenerator-aware algorithms alongside traditional routing and spectrum assignment (RSA) methods and leverages parallel processing to enhance simulation efficiency. While SimEON claims to be fully open-source, access to the code is currently limited, posing challenges for broader adoption in the research community.

Despite the advancements these simulators offer, their features have been considered but are not sufficient for comprehensive research needs. While numerous other simulators exist in the field, such as \cite{ieeeants} and \cite{ceons}, they were not highlighted due to their limited scope or inability to address the full range of functionalities required for advanced SDON research. While the simulators discussed above provide valuable tools for certain aspects of SDON research, they lack several critical features necessary for a truly comprehensive simulation environment. These include robust unit testing for ensuring simulation quality, detailed and standardized documentation for setup and assumptions, and support for high-performance computing clusters. Furthermore, essential functionalities like GitHub pipelines for quality control, standardized guidelines, multi-processing support, and configuration files are often incomplete or entirely absent. The absence of these features limits the applicability of these simulators for more rigorous research and highlights the need for a more robust, all-encompassing Flexible Unified Simulator for Intelligent Optical Networking (FUSION).

This gap underscores the necessity for a new simulator that not only integrates these missing functionalities but also sets a higher standard for the research community. By offering a fully open-source, high-quality platform that addresses the limitations of existing tools, this new simulator aims to drive innovation and support the growing demands of SDON research.

\section{System Architecture}
\label{sec:system_architecture}

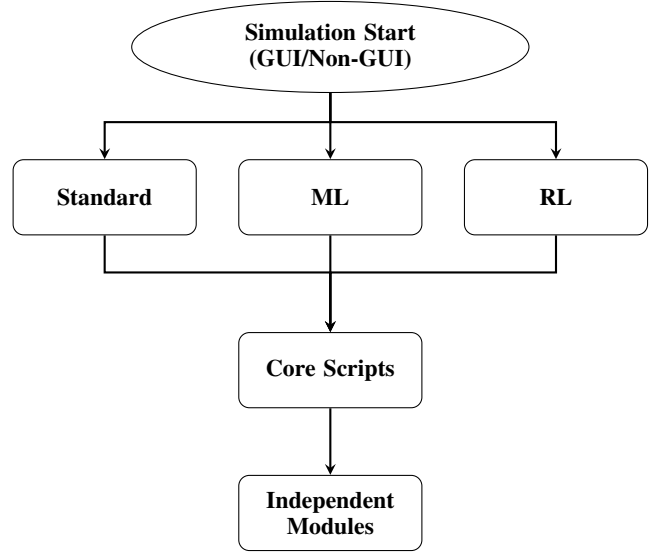
\begin{figure}[h]
    \centering
    \begin{tikzpicture}[
    node distance=2cm and 3cm, 
    every node/.style={font=\sffamily\small}, 
    align=center,
    process/.style={rectangle, draw, text width=2.2cm, minimum height=1cm, font=\bfseries\small, rounded corners}, 
    startstop/.style={ellipse, draw, text width=3.5cm, minimum height=1cm, font=\bfseries\small}, 
    arrow/.style={thick,->,>=stealth}
]

    \node (start) [startstop] {Simulation Start \\ (GUI/Non-GUI)};
    \node (std_sim) [process, below of=start, xshift=-3cm] {Standard};
    \node (ml_sim) [process, below of=start] {SL};
    \node (rl_sim) [process, below of=start, xshift=3cm] {RL};
    \node (src) [process, below of=ml_sim, yshift=-0.3cm] {Core Scripts};
    \node (modules) [process, below of=src, yshift=0.1cm] {Independent Modules};

    \draw [arrow] (start) -- ++(0,-1) -| (std_sim); 
    \draw [arrow] (start) -- ++(0,-1) -| (ml_sim); 
    \draw [arrow] (start) -- ++(0,-1) -| (rl_sim); 
    \draw [arrow] (std_sim) -- ++(0,-1) -| (src); 
    \draw [arrow] (ml_sim) -- ++(0,-1) -| (src); 
    \draw [arrow] (rl_sim) -- ++(0,-1) -| (src); 
    \draw [arrow] (src) -- (modules); 
\end{tikzpicture}
    \caption{High-level design of the FUSION architecture.}
    \label{fig:high_level_design}
\end{figure}

The system architecture of our Python-based SDON simulator (FUSION) is designed to be modular and flexible, accommodating a range of simulation types, including those driven by RL, supervised learning (SL), and traditional non-AI methods \cite{SDON_Simulator}. At the core of this architecture are the core scripts, which manage essential operations such as request generation, routing, spectrum assignment, and quality of transmission. These core components are coordinated by the central engine script, as illustrated in Figure~\ref{fig:high_level_design}.

\begin{figure}[h]
    \centering
    \fbox{\includegraphics[width=\columnwidth]{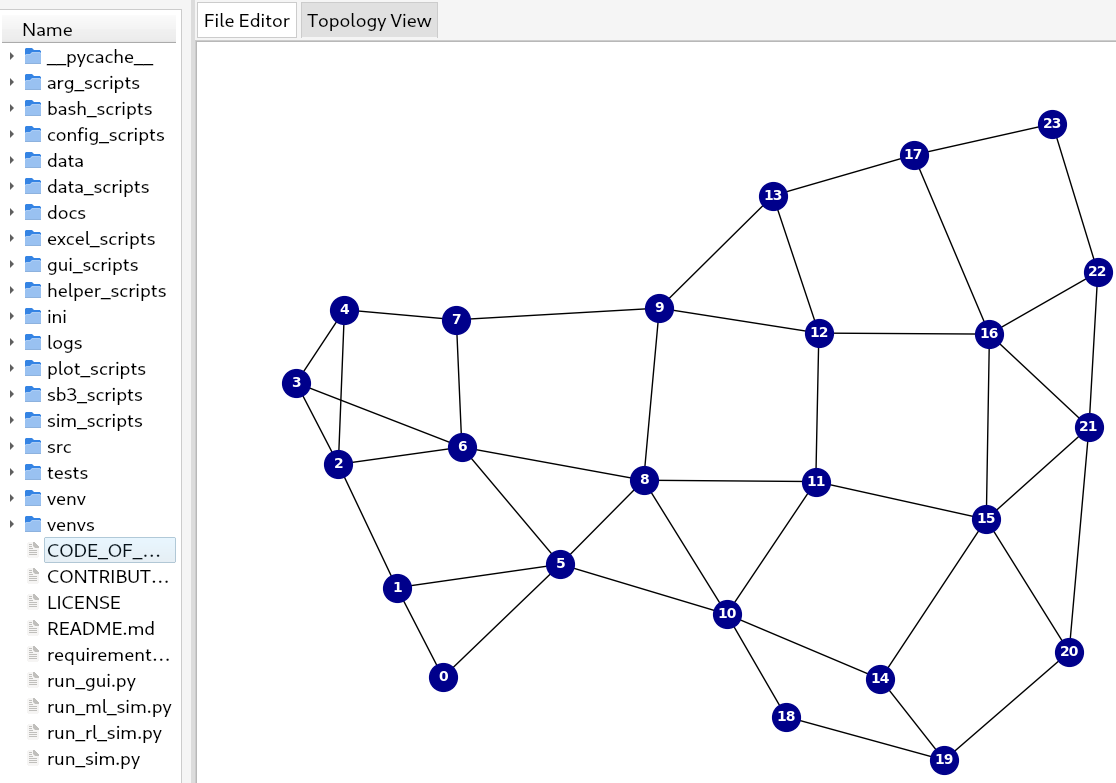}}
    \caption{FUSION GUI displaying the USNet Topology.}
    \label{fig:gui_screenshot}
\end{figure}

Users have the flexibility to run simulations either through a GUI or by directly utilizing configuration files. The GUI enhances usability by providing a text editor, the ability to view and select various topologies, configure simulation parameters, and execute simulations. A screenshot of the GUI is shown in Figure~\ref{fig:gui_screenshot}.

The helper scripts module plays a crucial role in organizing utility functions that are essential across the simulator, including statistical calculations and data export in formats like CSV, Excel, or JSON. For visual analysis, the plot scripts module generates detailed visual outputs from simulation data, assisting users in interpreting and analyzing results effectively. Additionally, the simulator integrates with advanced RL libraries such as Stable Baselines 3 (SB3), RLZoo3, and Gymnasium through the SB3 scripts module, enabling sophisticated AI-driven simulations.

Data management within the simulator is handled by the data scripts module, which oversees data generation and output, particularly to Excel. The configuration scripts module ensures that simulation input parameters are correctly structured and managed, whether running single or multi-processed simulations. For high-performance computing (HPC) environments, the bash scripts directory supports essential tasks such as creating environments, checking job priorities, submitting and managing jobs, and monitoring memory usage via a Slurm job scheduler. The argument scripts module centralizes all simulation run arguments, ensuring consistency and ease of configuration.

To maintain high quality and reliability, the simulator includes a robust tests module. This module is responsible for testing all other modules and scripts, ensuring that they function correctly and efficiently. We ensure 90\% unit testing coverage to guarantee the thoroughness of the tests, which are critical to maintaining the integrity of the simulator, especially when integrating new features or updates.

Documentation is automatically generated using Sphinx and is stored in the documents directory. This documentation includes comprehensive guides and references for users, ensuring they have the necessary information to effectively utilize the simulator. 

Table~\ref{table:modules_overview} provides an overview of the modules and directories within the FUSION simulator, summarizing their primary functionalities.

\begin{table}
\centering
\renewcommand{\arraystretch}{1.2} 
\begin{tabular}{|p{4cm}|p{4cm}|}  
\hline
\textbf{Module/Directory} & \textbf{Description} \\
\hline
\textit{Core Scripts} & Manages engine, routing, and SDN \\
\hline
\textit{Helper Scripts} & Utilities for stats, data export \\
\hline
\textit{Plot Scripts} & Visual output generation \\
\hline
\textit{SB3 Scripts} & Connects RL libraries \\
\hline
\textit{Tests} & Unit tests for components \\
\hline
\textit{Data Scripts} & Data generation \\
\hline
\textit{Statistical Helpers} & Exports stats (CSV, Excel, JSON) \\
\hline
\textit{Documents} & Sphinx documentation \\
\hline
\textit{Configuration Scripts} & Simulation input setup \\
\hline
\textit{Bash Scripts} & HPC tasks, job management \\
\hline
\textit{Argument Scripts} & Centralized simulation arguments \\
\hline
\textit{GitHub Workflows} & Continuous Integration automation \\
\hline
\textit{Miscellaneous Files} & Licenses, guidelines \\
\hline
\end{tabular}
\renewcommand{\arraystretch}{1.0} 

\caption{Overview of Modules and Directories}
\label{table:modules_overview}
\end{table}

The simulator is built using Python and leverages various libraries tailored to artificial intelligence and data processing, such as scikit-learn for SL and Stable Baselines 3 for RL. This technology stack facilitates the integration of sophisticated AI algorithms while maintaining high performance and scalability.

In summary, the system architecture of our simulator is designed to be modular, flexible, and extensible, ensuring that all components work together efficiently while allowing for the easy addition of new features. The core scripts handle the simulator's central operations, while auxiliary modules and directories add functionality and support, making the simulator a powerful tool for optical network research and development.

\section{Methodology}
\label{sec:methodology}

The simulation process in FUSION is highly configurable, allowing users to tailor the simulation environment to their specific needs. The configuration file provides a comprehensive interface to modify various parameters, particularly in relation to crosstalk-aware (XT-aware) algorithms, SL algorithms, and RL algorithms. The user can also select traditional routing and spectrum assignment algorithms, offering flexibility in how the simulations are conducted.

FUSION supports a variety of network topologies, allowing users to simulate different network environments. The topologies currently supported include the Dutch Telecom (DT), United States (US), Pan-European, and the National Science Foundation (NSF) networks. Traffic requests in the simulator are generated using random distributions: source and destination nodes are selected using a uniform distribution, while the arrival and holding times are modeled with an exponential distribution. These configurations provide a realistic simulation environment that can be tailored to explore a variety of scenarios.

\subsection{Algorithms and Techniques}

FUSION implements a diverse set of algorithms, each designed to address specific challenges in optical network management.

\subsubsection{Standard Routing and Spectrum Assignment Algorithms}
Users can select from traditional Routing and Spectrum Assignment (RSA) algorithms to simulate network scenarios under various conditions. For routing, the simulator supports algorithms such as k-shortest path (KSP) and shortest path (SP). For spectrum assignment, users can choose from first-fit (FF), best-fit (BF), and last-fit (LF) algorithms. These traditional algorithms provide a baseline for comparison against more advanced techniques, enabling users to explore different strategies for optimizing network performance.

\subsubsection{Crosstalk-Aware Algorithms}
The inter-core crosstalk (XT) aware algorithm in the simulator allows users to model real-world XT interference scenarios during spectrum assignment. By carefully selecting channels to minimize the impact of XT, users can simulate and evaluate the effects of XT on network performance. This feature is particularly useful for exploring how different levels of XT might affect the quality of transmission and for optimizing channel assignments to mitigate interference. The theoretical basis for this algorithm involves calculating potential XT levels and selecting routes that minimize these levels. The relevant theoretical equations for calculating XT for slot $s$ of core $c$ on path $P$ are as follows:

\begin{equation}
\label{eq:xtpath}
XT\left( P, c, s \right) = \sum\limits_{{e} \in {P}} \sum\limits_{{c^\prime} \in {A_{e,c,s}}} \frac{{1 - \exp \left( { - 2 \cdot h \cdot L\left( e \right)} \right)}}{{1 + \exp \left( { - 2 \cdot h \cdot L\left( e \right)} \right)}},
\end{equation}

where $L\left( e \right)$, $h$, and $A_{c,e}$ represent link length, power coupling coefficient, and the list of adjacent cores with active overlapping slots for slot $s$ of core $c$ on link $e$, respectively. This equation allows the simulator to quantify and minimize XT during resource allocation \cite{XTAR-lanman2024}.

\subsubsection{Reinforcement Learning Algorithms}
For RL, the simulator integrates a range of algorithms, including Q-learning, epsilon-greedy bandit, and Upper Confidence Bound (UCB) bandit. These RL algorithms can be applied to both routing and spectrum assignment decisions, allowing the simulator to dynamically optimize network performance.

The Q-learning algorithm is based on the following update equation:

\begin{equation}
    Q(s,a) \leftarrow Q(s,a) + \alpha \left[ r + \gamma \max_{a'} Q(s',a') - Q(s,a) \right]
\end{equation}

\noindent where \( Q(s,a) \) is the value of taking action \( a \) in state \( s \), \( \alpha \) is the learning rate, \( r \) is the reward, \( \gamma \) is the discount factor, \( s' \) is the next state that the agent transitions to after taking action \( a \), and \( \max_{a'} Q(s',a') \) represents the maximum Q-value over all possible actions \( a' \) in the next state \( s' \).

\noindent This equation governs how the Q-learning algorithm updates its knowledge based on the environment's feedback.

The epsilon-greedy bandit algorithm, which balances exploration and exploitation, follows this selection strategy:

\begin{equation}
    a =
    \begin{cases} 
    \text{random action}, & \text{with probability } \epsilon \\
    \max_{a} Q(s,a), & \text{with probability } 1 - \epsilon
    \end{cases}
\end{equation}

\noindent where \( \epsilon \) is the probability of choosing a random action, allowing the agent to explore the action space while also exploiting the current knowledge.

The Upper Confidence Bound (UCB) bandit algorithm is designed to select actions that maximize the upper confidence bound of the expected reward. The action selection strategy for UCB is given by:

\begin{equation}
    a_t = \arg\max_{a} \left[ \hat{\mu}_a + c \sqrt{\frac{2 \log t}{n_a}} \right]
\end{equation}

\noindent where \( \hat{\mu}_a \) is the estimated mean reward of action \( a \), \( c \) is a confidence level parameter, \( t \) is the current time step, and \( n_a \) is the number of times action \( a \) has been selected. This equation balances exploration and exploitation by considering both the expected reward and the uncertainty in the estimates.

In addition to these custom RL algorithms, FUSION includes support for all algorithms available in the Stable Baselines 3 library. This integration provides users with access to a wide range of state-of-the-art Deep-RL techniques, such as Asynchronous Advantage Actor Critic (A3C), Deep Q Network (DQN), and Proximal Policy Optimization (PPO), further enhancing the simulator's capability to optimize network performance in various scenarios.

\subsubsection{Supervised Learning Algorithms}
In addition to RL, the simulator includes SL algorithms such as the decision tree classifier. This algorithm can be used as a classifier for routing and spectrum assignment, enabling the simulation of different routing strategies based on classification of network states. The decision tree algorithm works by recursively partitioning the data space based on the features that provide the highest information gain or Gini impurity reduction. The algorithm’s decision-making process is governed by the following equation for information gain:

\begin{equation}
    IG(D, A) = H(D) - \sum_{v \in \text{values}(A)} \frac{|D_v|}{|D|} H(D_v)
\end{equation}

\noindent where \( IG(D, A) \) is the information gain of attribute \( A \) with respect to dataset \( D \), and \( H(D) \) is the entropy of the dataset \( D \). This equation determines how the decision tree splits the data at each node, which can be applied to optimize routing decisions based on network conditions.

Additionally, users can incorporate other supervised and unsupervised learning algorithms from the scikit-learn library if they wish to implement alternative or more complex classifiers. Examples of such algorithms include Support Vector Machines (SVM), Random Forest, and K-Nearest Neighbors (KNN). This flexibility allows users to explore a wide range of SL techniques for optimizing routing and spectrum assignment in FUSION.

\subsection{Comparison and Validation}

To ensure reliable and accurate simulations, FUSION includes extensive validation and testing procedures, with 90\% unit testing coverage across all modules. These tests focus on validating the critical algorithms and techniques, particularly when integrating new features or updates. The robust testing framework ensures the simulator functions as expected under various configurations and scenarios.

A key strength of the simulator lies in its extensive feature set, which surpasses that of other available simulators. Table~\ref{table:simulator_comparison} highlights these key features in comparison with other prominent optical network simulators, demonstrating that FUSION offers a broader range of capabilities. The simulator excels in areas critical to modern optical network research, such as crosstalk awareness, RL, support for multi-core optical fibers, multi-band support for spectrum assignment, and high-performance computing integration. These unique features make the simulator a powerful tool for both academic research and practical applications in the field of optical networks.

\begin{table*}
\centering
\renewcommand{\arraystretch}{1.2}
\begin{tabular}{|p{3.7cm}|>{\centering\arraybackslash}p{2.3cm}|>{\centering\arraybackslash}p{2.3cm}|>{\centering\arraybackslash}p{2.3cm}|>{\centering\arraybackslash}p{2.3cm}|>{\centering\arraybackslash}p{2.3cm}|}
\hline
\rowcolor{gray!20} \centering \textbf{Feature} & \textbf{FUSION} & \textbf{Optical RL-Gym} & \textbf{SimTON} & \textbf{SimEON} & \textbf{OMNeT++} \\
\hline

\rowcolor{gray!20} \textbf{Artificial Intelligence} & & & & & \\
\hline
Reinforcement Learning & X & &  &  & \\
\hline
Deep Reinforcement Learning & X & X &  &  &  \\
\hline
Supervised Learning & X &  &  &  & \\
\hline

\rowcolor{gray!20} \textbf{Network Modeling} & & & & & \\
\hline
Physical Impairment Modeling & X & X & X & X & X \\
\hline
Multi-Band Support & X & X &  &  & \\
\hline
Multi-Core Fibers & X & X &  &  & X \\
\hline
Multi-Processed Simulations & X & &  & X & X \\
\hline

\rowcolor{gray!20} \textbf{Usability and Development} & & & & & \\
\hline
Graphical User Interface & X &  & X &  & X \\
\hline
Unit Test Coverage & X & X & & & \\
\hline
Documentation and Tutorials & X & X &  &  & X \\
\hline
Continuous Integration Pipelines & X &  & & & \\
\hline

\rowcolor{gray!20} \textbf{Licensing and Output} & & & & & \\
\hline
Open-Source & X & X & X & X & X \\
\hline
Plotting & X & X & X & X & X \\
\hline
Custom Result Output & X &  &  &  & \\
\hline

\end{tabular}
\renewcommand{\arraystretch}{1.0}

\caption{Comparison of Features Between FUSION and Other Optical Network Simulators}
\label{table:simulator_comparison}
\vspace{-0.3cm}
\end{table*}

\section{Performance Evaluation}
\label{sec:performance_evaluation}

This section evaluates the performance of the proposed simulator by analyzing three key aspects: light segment slicing, XT-aware, and RL-based algorithms. The evaluations demonstrate the simulator's capability in reducing blocking probability and optimizing network performance. The results are presented through comparative analysis, highlighting the effectiveness of advanced slicing techniques and other innovative methods.

\subsection{Light Segment Slicing}

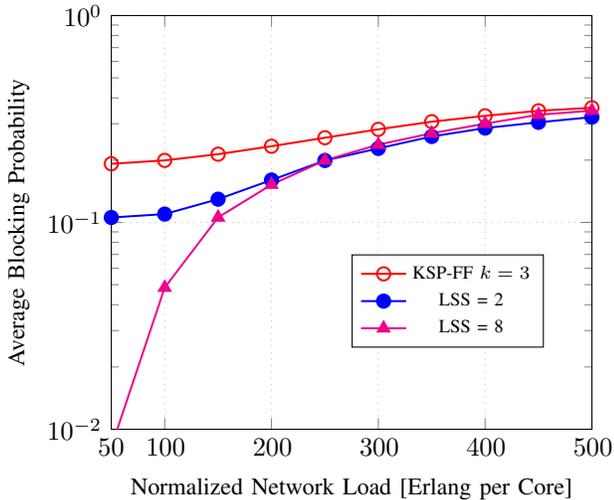
\begin{figure}
    \centering
    \begin{tikzpicture}[scale=1]
\begin{semilogyaxis}[
    xlabel={\small Normalized Network Load [Erlang per Core]},
    ylabel={\small Average Blocking Probability},
    xmin=50, xmax=500,
    ymin=1e-2, ymax=1,
    xtick={50, 100,200,300,400,500},
    x tick label style={align=center,text width=0.5cm},
    ytick={1e-2, 1e-1, 1},
    yticklabels={$10^{-2}$,$10^{-1}$,$10^{0}$},
    legend pos=north west,
    legend style={at={(0.5,0.31)}, scale=0.2, anchor=west},
    ymajorgrids=true,
    xmajorgrids=true,
    grid style=dotted,
    title style={at={(-0.18,-0.125)}, anchor=north, yshift=-0.1},
    title={\small},
    table/col sep=comma,
    width=0.9\linewidth,  
    height=0.8\linewidth, 
]
\addplot [red, mark=o, mark size=2.5pt, line width=0.75pt] table[y=blocking, x=erlang]{data/slicing/one_segments.csv};
\addplot [blue, mark=*, mark size=2.5pt, line width=0.75pt] table[y=blocking, x=erlang]{data/slicing/two_segments.csv};
\addplot [magenta, mark=triangle*, mark size=2.5pt, line width=0.75pt] table[y=blocking, x=erlang]{data/slicing/eight_segments.csv};

\legend{{\scriptsize KSP-FF $k = 3$}, {\scriptsize LSS = 2}, {\scriptsize LSS = 8},}

\end{semilogyaxis}
\end{tikzpicture}
    \caption{BP vs. Load for Light-Segment Slicing.}
    \label{fig:lss}
    \vspace{-0.5cm}
\end{figure}

The Light Segment Slicing (LSS) method introduces a segmentation mechanism that divides light paths into multiple segments, enhancing spectrum utilization and reducing blocking probability. Simulations were conducted on the USNet topology with 4 cores, allowing 1 (KSP-FF), 2, and 8 segments. Fig. \ref{fig:lss} shows that as the number of segments increases, blocking probability decreases significantly, particularly with 8 segments at earlier traffic volumes. This approach demonstrates a clear improvement in network performance, making LSS an effective strategy for optimizing optical networks. For more details, refer to \cite{yue_slice_one}.

\subsection{XT-Aware Algorithms}

\begin{figure}
    \centering
    \begin{tikzpicture}[scale=1]
\begin{semilogyaxis}[
    xlabel={\small Normalized Network Load [Erlang per Core]},
    ylabel={\small Average Blocking Probability},
    xmin=50, xmax=500,
    ymin=2e-4, ymax=1,
    xtick={100,200,300,400,500,600},
    x tick label style={align=center,text width=0.5cm},
    ytick={1e-5,1e-4,1e-3,1e-2,0.1,1},
    legend pos=north west,
    legend style={at={(0.5,0.31)}, scale=0.2, anchor=west},
    ymajorgrids=true,
    xmajorgrids=true,
    grid style=dotted,
    title style={at={(-0.18,-0.125)}, anchor=north, yshift=-0.1},
    title={\small},
    table/col sep=comma,
    width=0.9\linewidth,  
    height=0.8\linewidth, 
]
\addplot [red, mark=o, mark size=2.5pt, line width=0.75pt] table[y=blocking]{data/xt_aware/KSP1/KSP1_first_fit_1.csv};
\addplot [blue, mark=o, mark size=2.5pt, line width=0.75pt] table[y=blocking]{data/xt_aware/KSP3/KSP3_first_fit_1.csv};
\addplot [red, mark=*, mark size=2.5pt, line width=0.75pt] table[y=blocking]{data/xt_aware/KSP1/KSP1_prioritized_first_fit_8.csv};
\addplot [blue, mark=*, mark size=2.5pt, line width=0.75pt] table[y=blocking]{data/xt_aware/KSP3/KSP3_prioritized_first_fit_8.csv};
\addplot [magenta, mark=square, mark size=2.5pt, line width=0.75pt] table[y=blocking]{data/xt_aware/beta_1e-06/beta_1e-06_prioritized_first_fit_8.csv};

\legend{{\scriptsize 1SP-FF}, {\scriptsize 3SP-FF}, {\scriptsize SP-XTA-SMCSA-8}, {\scriptsize 3SP-XTA-SMCSA-8}, {\scriptsize XTAR-$\pi_1$(0)}, {\scriptsize $\pi_1$(0.1)}, {\scriptsize $\pi_1$(0)}, {\scriptsize $\pi_2$}}

\end{semilogyaxis}
\end{tikzpicture}
    \caption{BP vs. Load for XTAR \& XTA-SMCSA.}
    \label{fig:xt_aware}
    \vspace{-0.5cm}
\end{figure}
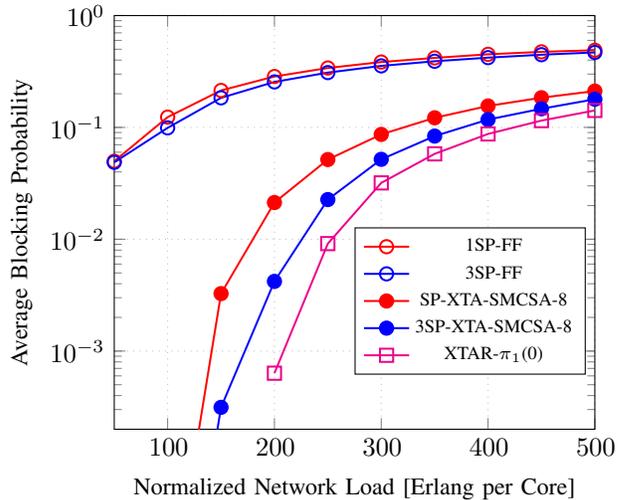

The XT-aware bandwidth-slicing modulation, core, and spectrum assignment (XTA-SMCSA) method integrates a light-segment slicing approach with the prioritized core and first-fit spectrum allocation algorithms, along with precise link-based XT calculations, to overcome the fragmentation issue, XT, and amplified spontaneous emission (ASE) noises simultaneously. Fig. \ref{fig:xt_aware} illustrates the request blocking probability of the XT-aware routing (XTAR) and KSP ($k$ = 1 (1SP) and $k$ = 3 (3SP)) routing algorithms along with XTA-SMCSA method and a maximum allowed slices of 8 compared to 1SP and 3SP with first fit core and spectrum allocation (1SP-FF and 3SP-FF) with precise link-based XT calculation versus normalized network load. These previously published results, obtained using the simulator, demonstrate that the implementation of the slicing approach significantly reduces the blocking probability, resulting in no requests being blocked at 50 Erlang. Additionally, incorporating XTAR with XTA-SMCSA, considering Policy I for \(\alpha=0\) (\(\pi_1(0)\)), outperforms both other approaches, completely eliminating the blocking probability up to 200 Erlangs. For more details, refer to \cite{XTAR-lanman2024}.

\subsection{Reinforcement Learning-Based Path Selection}

\begin{figure}
    \centering
    \begin{tikzpicture}[scale=1]
\begin{axis}[
    xlabel={\small Episodes [Simulation Runs]},
    ylabel={\small Average Blocking Probability},
    grid=major,
    grid style=dotted,
    xmin=-1, xmax=100,
    ymin=0, ymax=0.2,
    xtick={0, 20, 40, 60, 80, 100},
    ytick={0.05, 0.10, 0.15},
    legend style={at={(0.98,0.85)}, anchor=north east, font=\scriptsize},
    legend cell align={left},
    yticklabel style={
        /pgf/number format/fixed,
        /pgf/number format/fixed zerofill,
        /pgf/number format/precision=3,
    },
    width=0.9\linewidth,  
    height=0.8\linewidth, 
    smooth,
]

\addplot[
    smooth,
    solid,
    color=blue,
    line width=0.9pt,
] table [x=episodes, y=q_learning, col sep=comma] {data/reinforcement_learning/best_performing.csv};
\addlegendentry{\scriptsize Q-Learning}

\addplot[
    smooth,
    solid,
    color=red,
    line width=0.9pt,
] table [x=episodes, y=epsilon_greedy_bandit, col sep=comma] {data/reinforcement_learning/best_performing.csv};
\addlegendentry{\scriptsize Epsilon-Greedy Bandit}

\addplot[
    smooth,
    solid,
    color=gray,
    line width=0.9pt,
] table [x=episodes, y=ucb_bandit, col sep=comma] {data/reinforcement_learning/best_performing.csv};
\addlegendentry{\scriptsize UCB Bandit}

\addplot[
    smooth,
    dashed,
    color=magenta,
    line width=0.9pt,
] table [x=episodes, y=spf, col sep=comma] {data/reinforcement_learning/best_performing.csv};
\addlegendentry{\scriptsize SPF-FF}

\addplot[
    smooth,
    dashed,
    color=orange,
    line width=0.9pt,
] table [x=episodes, y=ksp, col sep=comma] {data/reinforcement_learning/best_performing.csv};
\addlegendentry{\scriptsize KSP-FF $k=4$}

\end{axis}
\end{tikzpicture}
    \caption{BP vs. Episodes for 750 Erlang.}
    \label{fig:path_agents}
    \vspace{-0.3cm}
\end{figure}
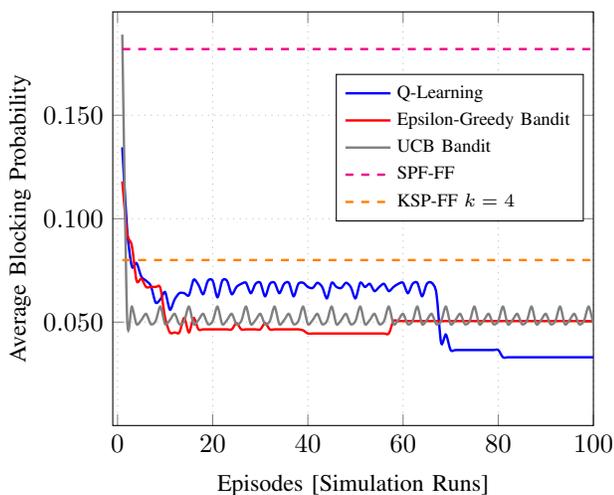

The path selection performance of different RL algorithms, namely Q-learning, epsilon-greedy bandit, and UCB bandit, is evaluated against traditional routing methods such as SPF-FF and KSP-FF (with $k=4$). Fig. \ref{fig:path_agents} presents the average blocking probability versus episodes for these algorithms under a network load of 750 Erlangs. The results indicate that R-based methods significantly outperform traditional algorithms in reducing blocking probability, particularly in the latter episodes where the algorithms have learned effective routing strategies. Q-learning, in particular, demonstrates the lowest blocking probability across all methods, capitalizing on its ability to dynamically adapt to varying network conditions. The epsilon-greedy bandit also shows robust performance, especially in the initial episodes, due to its balanced exploration and exploitation strategy.

\section{Conclusion}
\label{sec:conclusion}


FUSION is a comprehensive, open-source optical network simulator that addresses gaps in existing tools by supporting both traditional and advanced machine learning and RL techniques. It enables researchers to explore and optimize a wide range of network scenarios with a user-friendly interface and thorough testing framework. Performance evaluations show FUSION's accuracy in modeling complex SDON environments, highlighting its potential to drive innovation in optical network research. Future work will expand its capabilities, including new machine learning models and multi-band/multi-core fiber support for emerging technologies.

\bibliographystyle{IEEEtran}
\bibliography{references.bib}

\vspace{12pt}

\end{document}